\documentclass[aps, twocolumn, showpacs,groupedaddress]{revtex4}
\usepackage{amsmath}
\usepackage[dvips]{graphicx}
\usepackage[]{caption}
\usepackage{epsfig}
 
\begin{document}
\bibliographystyle{apsrev}

\title{Sedimentation profiles of systems with reentrant melting behavior} 

\author{J.~Dzubiella}
\email[e-mail address: ] {joachim@thphy.uni-duesseldorf.de}
\affiliation{Institut f{\"u}r Theoretische Physik II,
  Heinrich-Heine-Universit\"at D{\"u}sseldorf, 
  Universit\"atsstra{\ss}e 1, D-40225 D\"usseldorf, Germany}
\author{H.~M.~Harreis}
\affiliation{Institut f{\"u}r Theoretische Physik II,
  Heinrich-Heine-Universit\"at D{\"u}sseldorf, 
  Universit\"atsstra{\ss}e 1, D-40225 D\"usseldorf, Germany}
 \author{C.~N.~Likos}
\affiliation{Institut f{\"u}r Theoretische Physik II,
  Heinrich-Heine-Universit\"at D{\"u}sseldorf, 
  Universit\"atsstra{\ss}e 1, D-40225 D\"usseldorf, Germany}
 \author{H.~L\"owen}
\affiliation{Institut f{\"u}r Theoretische Physik II,
  Heinrich-Heine-Universit\"at D{\"u}sseldorf, 
  Universit\"atsstra{\ss}e 1, D-40225 D\"usseldorf, Germany}
\date{\today, submitted to Phys.\ Rev.\ E} 

\begin{abstract}
  We examine sedimentation density profiles of star polymer solutions
  as an example of colloidal systems in sedimentation equilibrium
  which exhibit reentrant melting in their bulk phase diagram. Phase
  transitions between a fluid and a fluid with an intercalated solid
  are observed below a critical gravitational strength $\alpha^{*}$.
  Characteristics of the two fluid-solid interfaces in the density
  profiles occurring in Monte Carlo simulations for $\alpha <
  \alpha^{*}$ are in agreement with scaling laws put forth in the
  framework of a phenomenological theory. Furthermore we detect
  density oscillations at the fluid-gas interface at high
  altitudes for high gravitational fields, which are verified with
  density functional theory and should be observable in surface
  scattering experiments.   
  
\end{abstract}

\pacs{64.70.-p, 61.25.Hq, 82.70.Dd}

\maketitle

\section{Introduction}

Colloidal particles in a suspension under gravitational influence show
spatial inhomogeneities due to the symmetry breaking induced by the
gravitational field. The problem of sedimentation of particles in the
presence of gravity has been of long scientific interest. The simplest
approximation is the one of non-interacting particles, valid in the
limit of dilute solutions. This approach leads to an exponential
sedimentation density profile, which was observed by Perrin for a
calculation of Boltzmann's constant in 1910 \cite{Perrin:1910}. Taking
into account particle interactions at higher concentrations will yield
corrections to the exponential density profile. For very small
gravitational strength, a local-density-approximation (LDA)
of density functional theory (DFT) is justified
\cite{Biben:etal:JCP:98:7330:1993,reviewHL}. In this case,
there is a one-to-one correspondence between the sedimentation
density profile  and the isothermal equation of state. This fact was
exploited to extract the hard sphere equation of state experimentally
by investigating  sterically stabilized colloids \cite{Piazza}.
Furthermore, within the LDA, a change in the
height $z$ corresponds to a local change of the chemical
potential $\mu$ of the bulk system.
This implies that, in the limit of small gravity,
the phase behavior becomes visible
as a function of height $z$, a feature
which was also been exploited to estimate the
hard sphere freezing transition \cite{Piazza}.
Surprisingly, comparison with Monte-Carlo (MC) simulations show
that the LDA is even reliable for relatively strong inhomogeneities
or gravitational strengths \cite{Biben:etal:JCP:98:7330:1993}.
This was further confirmed by comparing LDA against the exactly soluble
hard rod model in one spatial dimension.
While the LDA yields a  monotonic decaying density profile
$\rho(z)$, a layering shows up
near the hard wall of the container bottom.
Even crystallization can be induced by the bottom wall \cite{Biben2}.
As shown recently \cite{Heni},
details of this surface-induced crystallization may be significantly
influenced by a periodic wall pattern.
Indeed, pure colloidal crystals can be grown from sedimentation on a
patterned substrate \cite{vanBlaaderen1,vanBlaaderen2,vanBlaaderen3}.
In this case the gravitational field acts as an external force
enforcing and accelerating heterogenous nucleation and growth.
Other fascinating  phenomena in a gravitational field
relevant for colloidal suspensions are
phase transitions such as wetting  \cite{deGennes},
surface melting \cite{Beier}, as well as dynamical effects
as shock like fronts \cite{Snabre}, 
metastable phase formation \cite{Bocquet},
long-range velocity correlations
\cite{Segre}, stratification \cite{Mueth}, and crystal 
growth \cite{ackerson}.

While the equilibrium sedimentation of hard sphere suspensions
is well-understood
\cite{Biben:etal:JCP:98:7330:1993,Biben2,Piazza,granular,Levin:physicaA:287:100:2000},
charged suspensions are much more subtle
as they reveal an apparent mass which is smaller than the bare
mass at least for intermediate heights \cite{Piazza,Biben3,Loewen98,Simonin}.
In this paper, we study a third kind of effective interaction between
colloids, namely a very soft core as realized for star polymer
solutions \cite{christos:habil}. 
The qualitative new feature of those solutions as compared to the
traditional hard-sphere and Yukawa  interactions is that their phase diagram
exhibits a {\it reentrant melting behavior} for increasing density
\cite{Watzlaweck:etal:PRL:82:5289:1999}.
In fact, our analysis holds for any system with a reentrant melting
behavior but we will mainly focus explicitly on star polymers.
Star polymers consist of $f$ linear polymer arms
attached to a central common core. The complete bulk phase diagram for
star polymers in a good solvent was calculated in
\cite{Watzlaweck:etal:PRL:82:5289:1999} and exhibits several unusual
solid lattices as well as reentrant melting. As will be discussed in
detail in the following sections, due to the reentrant melting
behavior, unusual density profiles, featuring interesting effects,
arise and a wealth of scaling laws can be established.
  
The paper is organized as follows: In Sec.\
\ref{sec:Computer_Simulation} results of computer simulations of a
system of star polymers, interacting by means of an ultrasoft pair
potential \cite{Likos:etal:prl:80:4450:1998} are presented. In
Sec.\ \ref{sec:Theory}, we present a phenomenological theory giving
account of the sedimentation profiles observed in the computer
simulations.  Scaling laws are put forth. Also in Sec.\
\ref{sec:Theory}, density functional theory in a simplified hybrid weighted
density approximation (HWDA) is used to reproduce density oscillations
at the fluid-gas interface found in the simulation data. Concluding
remarks are contained in Sec.\ \ref{sec:Conclusion}.

\section{Computer Simulation}
\label{sec:Computer_Simulation}
\begin{center}
  \begin{figure}[htb]
    \epsfig{file=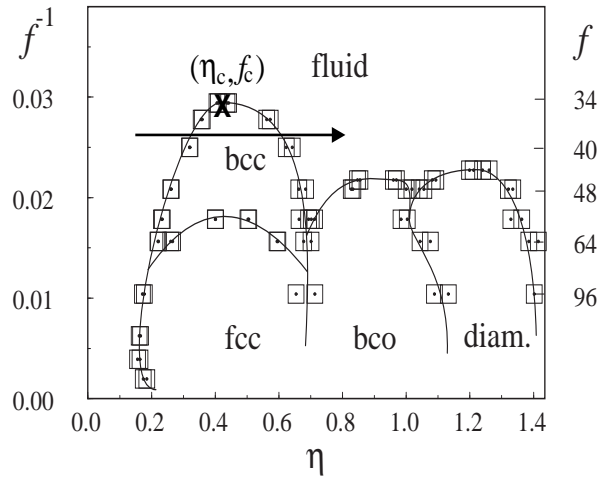, width=8cm, height=6.5cm, angle=0}
    \caption{Bulk phase diagram of star polymers interacting with
      potential (\ref{eq:star_star_potential}), calculated in
      \cite{Watzlaweck:etal:PRL:82:5289:1999}. Arm number $f$ is 
      plotted versus packing fraction $\eta$. The squares indicate the
      phase boundaries; solid lines are guide to the eye. The black
      cross denotes the point with critical arm number $f_{\rm c}\simeq
      34$ and corresponding density $\eta_{\rm c}\simeq 0.43$. The system
      is always fluid for arm numbers smaller than the critical arm
      number $f_{\rm c}$ and shows reentrant melting behavior
      for arm numbers $f_{\rm c}<f\lesssim 54$. 
      The arrow indicates a path
      through the phase diagram that is equivalent 
      with a change in the altitude $z$
      within the LDA for small $\alpha$. The four observed solid
      phases are body-centered cubic (bcc), face-centered cubic (fcc),
      body-centered orthogonal (bco), and diamond (diam.).} 
  \label{fig:bulk_phase_diagram}
\end{figure}
\end{center}

We performed canonical MC computer simulations keeping particle number
$N$, volume $V$, and temperature $T$ constant.  We used a simulation
box with squared periodic boundary conditions in $x$, $y$-direction
and semi-infinite geometry in $z$-direction where the particles were
confined only by the gravitational field for $z > 0$. The bottom wall
at $z =0$ was hard and interacting with the star polymers by means of
an effective star polymer-wall potential which is derived from the
effective star polymer-hard sphere interaction in the limit of a
sphere with zero curvature. The calculation was performed in
\cite{Jusufi:etal:tobepublished:2000}. It is of the following form:
\begin{eqnarray}
  \label{eq:star_wall_potential}
  \beta V_{\rm sw}(z) &= &\Lambda f^{3/2}\nonumber\\
  &\times&\begin{cases}
    \infty & {z<0} \\
     \xi_{2} -
    \ln({\frac{2z}{\sigma}})-(\frac{4z^{2}}{\sigma^{2}}-1)
    (\xi_{1}-\frac{1}{2})    &z < \frac{\sigma}{2}\\
    \xi_{2}(1-\text{erf}(2 \kappa
    z))/(1-\text{erf}(\kappa\sigma)) &\text{else.}
  \end{cases}\nonumber\\
\end{eqnarray}
With $z$ we denote the distance from the center of one star polymer to
the surface of the flat wall. $\sigma$ defines the so-called corona
diameter of a star polymer, which is related to its diameter of
gyration $\sigma_{\rm g}$ through $\sigma\simeq 0.66\sigma_{\rm g}$,
see \cite{Jusufi:etal:tobepublished:2000}.  The constants are
$\Lambda=0.24$, $\kappa{\sigma}=0.84$,
$\xi_{1}=1/(1+2\kappa^{2}{\sigma}^{2})$, $\xi_{2}=\frac{\sqrt{\pi}
  \xi_{1}}{\kappa\sigma}\exp(\kappa^{2}{\sigma}^{2})
(1-\text{erf}(\kappa{\sigma}))$ and the inverse thermal energy
$\beta=1/k_{B}T$. We emphasize that the range of the star-wall
interaction is of the order of one or two corona diameters, so that
the behavior of the sedimentation profiles for larger distances is not
influenced.  The star polymer pair potential is ultrasoft and is
described by the following equation
\cite{Likos:etal:prl:80:4450:1998}:
\begin{eqnarray}
  \label{eq:star_star_potential}
  \beta V_{\rm ss}(r) &= &\frac{5}{18} f^{3/2} \begin{cases}
     -\ln(\frac{r}{\sigma})+\frac{1}{1+ \sqrt{f}/2}
    &r<\sigma\\
    \frac{\sigma/r}{1+ \sqrt{f}/2}\exp(-
    \frac{\sqrt{f}}{2\sigma}(r-\sigma)) &\text{else,}
  \end{cases}\nonumber\\
\end{eqnarray}
with center-to-center distance $r$.  Both interactions are purely
entropic, hence they scale linearly with
temperature.  Previous work
\cite{Watzlaweck:etal:PRL:82:5289:1999} showed that a system of star
polymers interacting by means of the potential
(\ref{eq:star_star_potential}) possess a very rich and interesting
bulk phase diagram, see Fig.\ \ref{fig:bulk_phase_diagram}, exhibiting
reentrant melting and reentrant freezing transitions for arm numbers
$f_{\rm c} \lesssim f \lesssim 54$, with the critical arm number
$f_{\rm c}=34$.  As we will discuss in more detail below, it is the
reentrant melting that makes this type of system appropriate for the
analysis presented in this article.  The suspending liquid is assumed
to be incompressible. Furthermore we treat the solvent to be
continuous, neglecting possible effects of the discreteness of the
solvent particles. Given the size of the colloidal particles under
observation, the star polymers, this is a reasonable assumption.  In
the simulation, the initial configuration of the system was chosen to
be a body-centered cubic (bcc) solid to facilitate equilibration. Its
lattice constant $a$ 
was determined from a bulk system with a packing fraction
$\eta=\frac{\pi}{6}\rho\sigma^{3}\simeq 
0.5$ lying in the bcc-regime in the bulk phase diagram, see 
Fig.\ \ref{fig:bulk_phase_diagram}. The lateral box-dimensions were
chosen to be multiples of the lattice constant $a$.

The total number of particles was then fixed by prescribing a certain
value of the thermodynamic variable $\tau$, giving the number density
per unit surface. The density profile $\rho(z)$ is normalized as
\begin{equation}
  \label{eq:tau}
  \tau = \int_{0}^{\infty} \rho(z) dz.
\end{equation}
$\tau\sigma^{2}$ is the number of particles piled up over the area
$\sigma^{2}$ of the bottom wall. Typical system sizes were $N=2000$
particles and the Monte Carlo runs were extended over $N_{\rm MC}
\approx 500\,000$ cycles, each cycle comprising one trial move for each
of the $N$ particles. Besides the aforementioned thermodynamic
variable $\tau$, two further parameters characterize the state of the
system: First, the arm number $f$ of the star polymers, being the
number of polymer chains grafted on the central core. Second, the
dimensionless gravitational strength (or Peclet number)
\begin{equation}
  \label{eq:gravitational_parameter}
  \alpha = \frac{m g \sigma}{k_{B} T},
\end{equation}
which describes the ratio of the potential energy gain to the thermal
energy $k_{B}T$ for a particle of mass $m$, displaced by $\sigma$
in height in an external field with acceleration $g$. The three
parameters $f$, $\tau$ and $\alpha$ were varied over a broad range of
values. The particles were moved by employing the standard Metropolis algorithm.
\begin{center}
\begin{figure}[htb]
  \epsfig{file=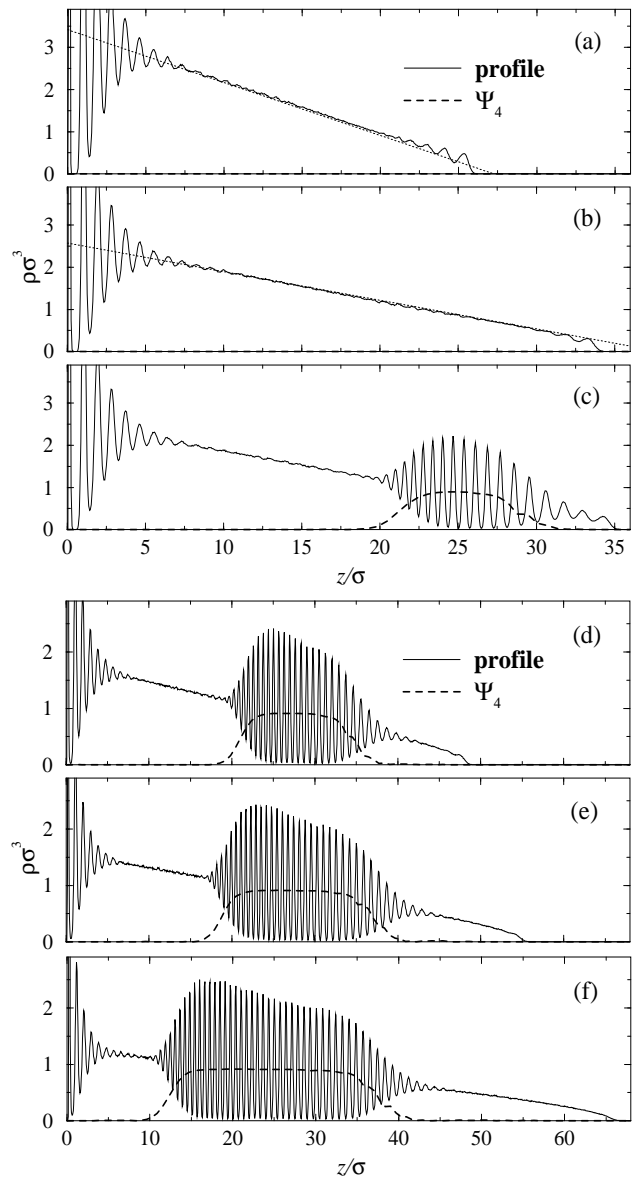, width=8.5cm, height=16.0cm, angle=0}
\caption{Sedimentation profiles of star polymers for an arm number
  $f=39$ and a density $\tau\sigma^{2}=48.87$. The gravitational
  strength $\alpha$ is decreased from (a) to (f) with (a)
  $\alpha=30.0$, (b) $\alpha=17.0$, (c) $\alpha=16.0$, (d)
  $\alpha=8.0$, (e) $\alpha=6.0$ and (f) $\alpha=4.0$. In plots (c) -
  (f) the order parameter $\Psi_{4}$ is also shown (dashed line) using
  the same $y$-scale as the profiles. In (a) and (b) a 
  straight line whose equation is derived within the LDA
  [see Eq.\ (\ref{eq:profile_in_tau})] is superimposed on
  the plots (dotted line).}
\label{fig:profiles}
\end{figure}
\end{center}

In Fig.\ \ref{fig:profiles} we show results for different gravitational
strengths $\alpha$, while $\tau\sigma^{2}=48.87$ and $f=39$ are
fixed. The $f=39$ star polymer system displays reentrant melting
in the fluid $\to$ bcc $\to$ fluid sequence, as seen along the arrow
in  
Fig.\ \ref{fig:bulk_phase_diagram}. The gravitational field
forces the local density $\sigma^3\rho(z)$ to take values that
scan the range from $\sigma^3\rho(z) = 0$ up to high values,
$\sigma^3\rho(z) \cong 3$. Thus, the local density `crosses through'
the range of the phase diagram where
the system displays a bulk bcc phase. It it intuitively expected
that the system will then feature a solid regime (for intermediate
densities) intercalated between two fluid regimes, at low and high
densities. We have found that this is indeed what happens but
{\it provided} that the gravitational strength does {\it not}
exceed a critical value $\alpha^{*}$, as we discuss below.

Let us start from the case where no solid phase appears.
For $\alpha > \alpha^{*}$ [Figs.\ \ref{fig:profiles}(a) and (b)],
we obtain density
profiles $\rho(z)$ that show three distinct features: First, there is
layering on the wall due to packing effects, typically extending over
several layers. As $z$ increases a fluid regime with density decaying
as a linear function of altitude $z$ can be distinguished. At some
height ($z\simeq 25\sigma$ in (a)) the density rapidly decays to zero.
At 
this strong inhomogeneity oscillations in density with wavelength
$\sigma$ can be distinguished in the sedimentation profile, which is
smooth elsewhere in the linear regime. The linear dependence of the
density profile on $z$, can be understood in terms of a local 
density functional mean-field
theory, as will be shown in Sec.\ \ref{sec:Phenomenological_Theory};
the corresponding results from 
this theory are shown in Figs.\ \ref{fig:profiles}(a) and
(b) with dotted lines.  
The density oscillations observed in the simulations were reproducible
in the framework of density functional theory using a simplified form
of the HWDA, as will be discussed in further detail in Sec.\ \ref{sec:WDA}.

\begin{center}
  \begin{figure}[htb]
  \epsfig{file=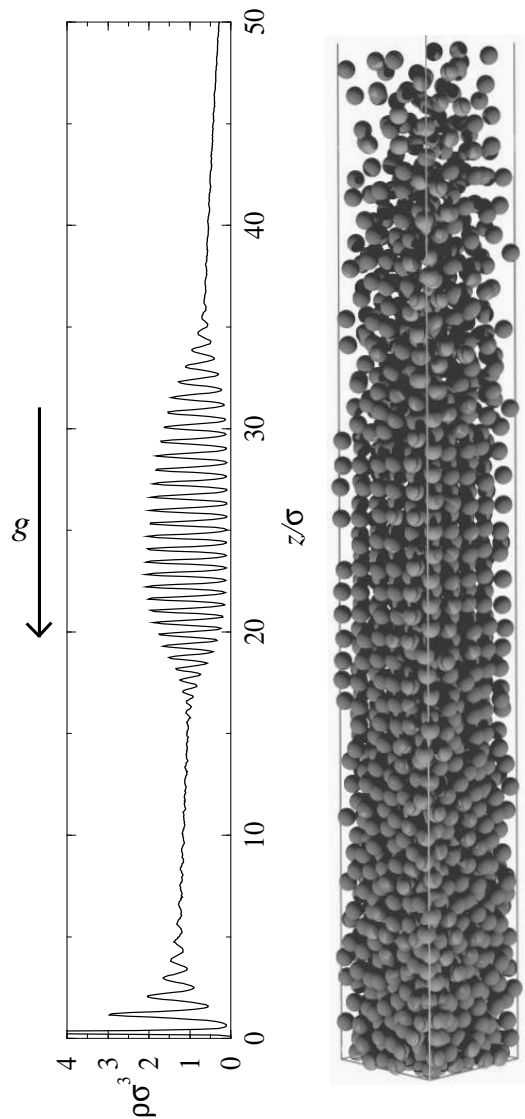, width=7cm, height=15cm, angle=0}
  \caption{Snapshot from MC simulation (right) shown with the
      corresponding equilibrium density profile (left). The star
      polymers are rendered as spheres with diameter $\sigma$. The
      parameters are: $f=39$, $\alpha=5.0$,
      $\tau\sigma^{2}=42.1$.}
    \label{fig:snapshot}
\end{figure}
\end{center}

By lowering the gravitational strength $\alpha$ further, a critical
strength
$\alpha^{*}$ in the range $16.0<\alpha^{*}<17.0$ is discovered. Below
$\alpha^{*}$ the density profiles qualitatively change and exhibit a
new feature. Strong density oscillations appear, a clear indication
for a crystalline phase. They extend over $10$ to $20$ star diameters,
equivalent to several crystalline layers. The length of the crystal
grows, as $\alpha$ decreases. A typical simulation snapshot is shown
in Fig.\ \ref{fig:snapshot} next to the corresponding equilibrium
density profile. Here, the well-ordered crystal phase in the middle of
the simulation box ($20\sigma \lesssim z\lesssim 30\sigma$) is clearly
visible.
 
As an additional check for crystalline order, we calculate the local
order parameter, $\Psi_{4}$, that checks for fourfold symmetry in two
dimensions around a given particle. It is defined by
\begin{equation}
  \label{eq:psi_4}
  \Psi_{4}(z) = \left|\left<\frac{1}{4N_{\rm l}}
      \sum_{j=1}^{N_{\rm l}}\sum_{<k>} e^{4 i
      \phi_{jk}}\right>\right|,
\end{equation}
where the $k$-sum includes the four nearest neighbors of the given
particle and the $j$-sum extends over $N_{\rm l}$ particles in the
corresponding layer. A layer is defined by a slab of thickness
$\delta\simeq 0.2\;a$, centered around the given particle at elevation
$z$, which is 
motivated by the `Lindemann melting rule', assuming a maximum particle
displacement of approximately $10\%$ around the equilibrium position
in a possible crystal regime.
The angular brackets indicate a canonical ensemble average.
$\phi_{jk}$ is the polar angle of the interparticle distance vector
with respect to a fixed reference frame. For ideal fourfold symmetry,
i.e., for a particle contained in a bcc-solid layer, $\Psi_{4} = 1$.
Due to thermal motion, small defects of the perfect crystalline
symmetry arise and usually values of $\Psi_{4} > 0.8$
\cite{Heni} are taken to be conclusive evidence
for a crystalline phase with fourfold-in-layer-symmetry. As can be
seen in Figs.\ \ref{fig:profiles}(c)-(f) our simulation data do
indeed show values up to $\Psi_{4} \approx 0.95$ in the region of the
density profile $\rho(z)$ which we already identified to be solid due
to the pronounced density oscillations.  

Comparing the interval of the
packing fraction in which crystallization occurs to the bulk phase
diagram in Fig.\ \ref{fig:bulk_phase_diagram}, we may thus conclude
that the intercalated solid regime is a manifestation of the reentrant
melting in the bulk phase diagram, mapped onto the $z$-axis in a
system under gravitational influence. The absence of freezing for
strong gravitational fields ($\alpha > \alpha^{*}$) can now be
at least qualitatively understood: for high values of $\alpha$,
the density profiles grow too fast as $z$ approaches the 
wall, so that the mapping
onto the $z$-axis results into a domain which is too narrow
to sustain crystalline order. In fact, as we will show in detail
in Sec.\ \ref{sec:Phenomenological_Theory}, a minimal, nonvanishing
thickness of the crystalline layer is necessary so that the latter
can be stably `nested' between the two fluid phases. 

\clearpage
\section{Theory}
\label{sec:Theory}

\subsection{Density Functional Theory in Local Density Approximation (LDA)}
\label{sec:LDA}
In order to predict scaling relations characterizing crystallization
in sedimentation profiles of star polymer solution we apply density 
functional theory within the framework of the 
local density approximation (LDA). The latter is a reliable theoretical
tool in cases where the density profile of the system varies slowly
with $z$, so that it can be considered as staying essentially 
constant at length scales set by the 
microscopic natural length of the system ($\sigma$ in this case). As can
be seen in Figs.\ \ref{fig:profiles}(a) and (b), this is indeed the
case if we discard the strong oscillations close to the wall (the
layering effect). As the range of these oscillations is much shorter
than the range of the density profile itself, the bulk of the free
energy of the system resides in the smooth ``ramp-like'' part of the
density profile and the use of the LDA is justified. Accordingly,
we will omit the star-wall potential from our considerations in this
subsection and consider only the effects of the 
external gravitational field $\Phi_{\rm ext}(z) = mgz$.

We work in the grand canonical ensemble and introduce the chemical 
potential $\mu$ and a variational grand potential per unit area,
$\tilde\Sigma(T,\mu;[\rho(z)])$ which is a functional of the density
profile. Introducing the ideal and excess per unit area contributions to the
intrinsic Helmholtz free energy of the system, $F_{\rm id}[\rho(z)]$
and $F_{\rm ex}[\rho(z)]$ respectively, we find that
in the LDA, the expression for $\tilde\Sigma(T,\mu;[\rho(z)])$ 
reads as:
\begin{eqnarray}
\nonumber
\tilde\Sigma(T, \mu, [\rho(z)]) & = & F_{\rm id}[\rho(z)] + F_{\rm ex}[\rho(z)] 
\\
\nonumber
  & & +\int {\rm d}z \Phi_{\rm ext}(z)\rho(z) - 
   \mu\int {\rm d}z \rho(z)
\\
\nonumber
& = &
k_BT\int_0^{\infty}{\rm d}z\rho(z)\left[\ln(\rho(z)\lambda^3) - 1\right] 
\\
\nonumber 
& + & \int_{0}^{\infty}{\rm d}z 
\left[f(\rho(z)) + (mgz-\mu)\rho(z)\right],
\\
& &
\label{dft.lda}
\end{eqnarray}
where $\lambda = \sqrt{h^2/2\pi m k_BT}$ is the thermal de Broglie
wavelength and
$f(\rho(z))$ is the Helmholtz free energy per unit
volume of the bulk fluid.
The minimization of $\tilde\Sigma$ with respect to $\rho(z)$ yields the
equilibrium profile $\rho_0(z)$; the value of the functional
at equilibrium, $\tilde\Sigma(T, \mu, [\rho_0(z)])$ is then the
Gibbs free energy per unit area, $\Sigma(T,\mu)$ of the system.
Setting $\delta\tilde\Sigma(T,\mu,[\rho(z)])/\delta\rho(z)|_{\rho_0(z)} = 0$ 
in Eq.\ (\ref{dft.lda}),
leads to: 
\begin{equation}
  \label{eq:equilibrium_profile}
 k_BT\ln(\rho_0(z)\sigma^3) + f'(\rho_0(z)) = \mu' -  mgz, 
\end{equation}
where $f'(x)$ denotes the derivative of $f(x)$ and
$\mu' = \mu - 3\ln(\lambda/\sigma)$ is a shifted chemical potential.

Due to the ultrasoft character of the logarithmic-Yukawa star-star
interaction $V_{\rm ss}(r)$, the star polymer system belongs to the class
of {\it mean-field fluids}
\cite{watz_struc98,lang:etal:gcm,likos:etal:exact,ard:mft:00,ard:royal:01},
for which the excess free energy density is  
a quadratic function of $\rho$, namely:
\begin{equation}
  \label{eq:free_energy_mean_field}
  f(\rho) \cong 
  \frac{\rho^{2}}{2}4\pi\int_{0}^{\infty}{\rm d}r\; r^{2} V_{\rm ss}(r) =:
  \frac{\hat V_{\rm ss}(0)\rho^2}{2},
\end{equation}
with the Fourier transform $\hat V_{\rm ss}(k)$ of the pair potential. 
This property is valid for high density
fluids provided their pair potential $V(r)$ is only slowly
diverging at the origin and decays fast enough to 
zero as $r \to \infty$, so that it is integrable. For
the more restrictive case of a nondiverging potential at $r = 0$, the
stronger condition $c(r) = -\beta V(r)$ 
holds 
approximately \cite{lang:etal:gcm,likos:etal:exact,ard:mft:00,ard:royal:01},
with $c(r)$ denoting the direct correlation function 
of the fluid \cite{hansen:mcdonald}. This gives 
again rise to Eq.\ (\ref{eq:free_energy_mean_field}) above
through the compressibility equation of state \cite{hansen:mcdonald}.

Using the dimensionless variables $x \equiv z/\sigma$, 
$\bar\rho(x) \equiv \rho(z)\sigma^3$, 
$B \equiv \beta\hat V_{\rm ss}(0)/\sigma^3$
and $\bar\mu \equiv \beta\mu'$ and introducing
Eq.\ (\ref{eq:free_energy_mean_field}) 
into Eq.\ (\ref{eq:equilibrium_profile}),
we obtain the equilibrium profile through the equation:
\begin{equation}
\ln(\bar\rho_0(x)) + B\bar\rho_0(x) = \bar\mu - \alpha x.
\label{eq.dimensionless}
\end{equation}
For star functionality $f = 39$ we obtain $B \cong 250$ and, for 
$f = 32$, $B \cong 204$ [see Eq.\ (\ref{eq:star_star_potential})].
Hence, the second term in the lhs of Eq.\ (\ref{eq.dimensionless}) above
dominates over the logarithmic term for densities 
$\bar\rho(x) \gtrsim 0.10$. As almost the entire simulation density 
profile fulfills this condition, we finally omit the logarithmic
term from Eq.\ (\ref{eq.dimensionless}) above and obtain thereby
a {\it linear} density profile:
\begin{equation}
  \label{eq:equilibrium_profile_linear_in_z}
  \bar\rho_0(x) = \begin{cases}
  0 & \text{for $x < 0$}, \\
  \frac{\bar\mu - \alpha x}{B} & \text{for $0 < x < \bar\mu/\alpha$},\\
  0 & \text{for $\bar\mu/\alpha < x$}.
  \end{cases}
\end{equation}
The chemical potential $\bar\mu$ is now determined through the 
normalization condition 
$\int_0^{\bar\mu/\alpha}{\rm d}x\bar\rho_0(x) = \tau\sigma^2 \equiv
\bar\tau$, yielding:
\begin{equation}
\bar\mu = \sqrt{2\alpha B \bar\tau},
\label{eq:chempot}
\end{equation}
and from Eq.\ (\ref{eq:equilibrium_profile_linear_in_z}) the final
expression for the density profile:
\begin{equation}
  \label{eq:profile_in_tau}
  \bar\rho_0(x) = \begin{cases}
  0 & \text{for $x < 0$}, \\
  \sqrt{\frac{2\alpha\bar\tau}{B}} - \frac{\alpha}{B}x & 
  \text{for $0 < x < \sqrt{\frac{2B\bar\tau}{\alpha}}$},
\\
  0 & \text{for $\sqrt{\frac{2B\bar\tau}{\alpha}} < x$}.
  \end{cases}
\end{equation}

The prediction 
(\ref{eq:profile_in_tau}) is compared against the MC
simulation results in Figs.\ \ref{fig:profiles}(a) and (b);
theory and simulation are in excellent
agreement. This linear dependence of the density profile on
$z$ is the first scaling prediction we make for such systems.
Moreover, by introducing Eq.\ (\ref{eq:equilibrium_profile_linear_in_z})
into Eq.\ (\ref{dft.lda}), and once more ignoring the logarithmic
term, we find that the Gibbs free energy per unit area 
$\Sigma(T,\mu)$ is a power-law of the chemical potential,
namely:
\begin{equation}
\beta\sigma^2\Sigma(T,\bar\mu) = -\frac{\bar\mu^3}{6\alpha B}.
\label{eq.gibbs}
\end{equation}
Accordingly, the Helmholtz free energy per unit area, 
$\beta\sigma^2F(T,\bar\tau) = \beta\sigma^2\Sigma(T,\bar\mu) + \bar\mu\bar\tau$ obeys
the scaling law:
\begin{equation}
\beta\sigma^2 F(T,\bar\tau) = \frac{2}{3}\sqrt{2\alpha B}\,{\bar\tau}^{3/2}.
\label{eq.helmholtz}
\end{equation}
The thermodynamic relation 
$\bar\mu = {\partial (\beta\sigma^2 F)}/{\partial \bar\tau}$
returns Eq.\ (\ref{eq:chempot}).

We now examine whether the density oscillations occurring at high
$z$-elevations, which are clearly visible in 
Figs.\ \ref{fig:profiles}(a), (b), can be obtained in the framework
of the full LDA, with the logarithmic term included, 
Eq.\ ({\ref{eq.dimensionless}). Though the latter is an implicit
equation for $\bar\rho(x)$, we do not need to solve it 
in order to answer the question at hand. The key observation is that
Eq.\ (\ref{eq.dimensionless}) delivers 
an explicit functional form
for the {\it inverse} function:
\begin{equation}
x(\bar\rho) = -\alpha^{-1}(\ln\bar\rho + B\bar\rho - \bar\mu).
\label{eq.inverse}
\end{equation}
If the LDA profile displayed oscillations, then 
$\bar\rho(x)$ would go through various maxima and
minima and there should be several points
$x_{\rm m}$ where the derivative $\bar\rho'(x_{\rm m})$ would vanish,
with the implication that the derivative of the inverse function,
$x'(\bar\rho_{\rm m})$, would {\it diverge} at the corresponding
density values $\bar\rho_{\rm m}$. From Eq.\ (\ref{eq.inverse})
above, we obtain $x'(\bar\rho) = -\alpha^{-1}(\bar\rho^{-1} + B) < 0$
for all $0 < \bar\rho < \infty$. The only divergence of 
$x'(\bar\rho)$ occurs for the trivial limit $\bar\rho \to 0$ and
corresponds to the exponential decay $\bar\rho(x) \propto e^{-\alpha x}$,
valid for high elevations. The LDA is incapable to reproduce this effect,
a feat that, in fact, could have been anticipated: these oscillations occur
at length scales $\sigma$, whereas the LDA is applicable when the
spatial inhomogeneity of the profile has a characteristic length
much larger than the latter. In Sec.\ \ref{sec:WDA},
we resort to a more powerful 
density functional approximation in order to reproduce this
feature of the density profile. 

\subsection{Phenomenological Landau Theory}
\label{sec:Phenomenological_Theory}
If one focuses close to the reentrant melting transition point and
studies the case for small gravity $(\alpha \ll 1)$, a
phenomenological Landau-like approach can be adopted to explore
further scaling predictions for the crystallization transition. We
study the situation sketched in Fig.
\ref{fig:sedimentation_profile_sketch}
\begin{figure}[hbt]
  \epsfig{file=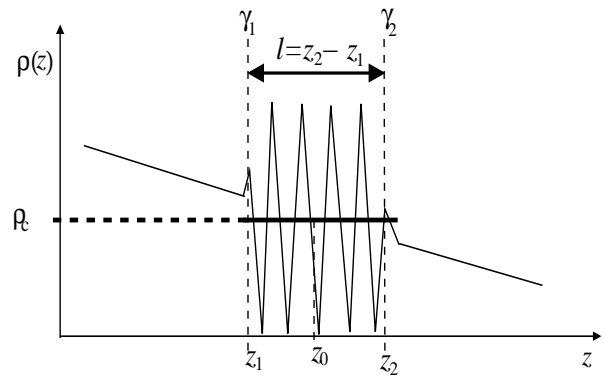, width=8cm, height=5cm, angle=0}
  \caption{Sketch of the situation in which 
   an intervening solid of width $l$ is nested between two fluids
   under the influence of a gravitational field.}
  \label{fig:sedimentation_profile_sketch}
\end{figure}
of a crystalline sheet of width $l$ intervening between two fluid
parts of the sedimentation profile. Let us define the excess grand
canonical free energy per unit area, $\Sigma_{\rm ex}(l)$, 
in such a situation with respect to a situation
where no crystallization takes place. This quantity is given by:
\begin{equation}
\Sigma_{\rm ex}(l) = \Sigma(l) - \Sigma(l = 0),
\label{eq:sigma_exc}
\end{equation}
with $\Sigma(l)$ being the grand canonical free energy per unit
area when a solid of thickness $l$ is present. Evidently, 
$\Sigma(l = 0)$ is the quantity given by Eq.\ (\ref{eq.gibbs}) above.
 
The excess grand canonical free
energy $\Sigma_{\rm ex}$ per unit area comprises of three parts:
\begin{enumerate}
\item The equilibrium surface tensions $\gamma_{1}$ and $\gamma_{2}$:
  These describe the additional free energy cost in creating the two
  solid-fluid interfaces at $z=z_{1}$ and $z=z_{2}$.
\item A thermodynamic contribution which essentially depends on the arm
  number. If $f > f_{\rm c}$ this contribution favors a solid sheet.
\item A free energy penalty due to an elastic distortion of the solid in 
  the external field.
\end{enumerate}
Hence:
\begin{equation}
  \label{eq:excess_grand_canonical_free_energy}
  \Sigma_{\rm ex} = \gamma_{1} + \gamma_{2} + \Sigma_{\rm TD} +
  \Sigma_{\rm elast}.
\end{equation}
Let us discuss the different contributions in more detail:

The surface tension will mainly control the relative orientation of
the solid with respect to the $z$-direction. One expects that a
close-packed surface of the bcc-solid (i.e. a (100) orientation) will
have smallest surface tension and will hence be the realized
orientation. In fact, this is what we found in our simulation data
presented in Figures \ref{fig:profiles} and \ref{fig:snapshot}. For
hard-sphere fcc-solids the 
interfacial fluid-solid 
free energy has been calculated recently in equilibrium by computer
simulation \cite{Laird}. Its order of magnitude is
\begin{equation}
  \label{eq:gamma_i_magnitude}
  \gamma_{i} \approx \frac{k_{B} T}{\sigma^{2}},\; \; (i = 1,2),
\end{equation}
where $\sigma$ is a microscopic length scale. 

The thermodynamic
contribution could be calculated within the LDA with $f(\rho, T)$
taken from liquid state theories for the fluid and solid cell theory
for the crystal. Here, we will simply focus on a Landau-type theory
close to the reentrant melting point characterized by a critical star
number density $\rho_{\rm c}=\frac{6}{\pi\sigma^{3}}\eta_{\rm c}$ and
the critical arm number 
$f_{\rm c}$, see Fig.\ \ref{fig:bulk_phase_diagram}. Performing a
Landau expansion and dropping the temperature 
dependence one gets
\begin{eqnarray}
  \label{eq:Landau_expansion_solid}
  f_{\rm s}(\rho) &= &f_{\rm s}(\rho_{\rm c}) + A_{\rm s} (f_{\rm c}-f)\nonumber\\
              &  &+f_{\rm s}'(\rho_{\rm c})(\rho-\rho_{\rm c})\nonumber\\
              &  &+\frac{1}{2} f_{\rm s}''(\rho_{\rm
              c})(\rho-\rho_{\rm c})^{2}+\dots\;.
\end{eqnarray}
Here, $f_{\rm s}(\rho)$ is the free energy per unit volume of the solid
phase and $A_{\rm s}$ is a constant governing the first leading term in an 
expansion around $f=f_{\rm c}$. 
Likewise in the fluid phase one has 
\begin{eqnarray}
  \label{eq:Landau_expansion_fluid}
  f_{\rm f}(\rho) &= &f_{\rm f}(\rho_{\rm c}) + A_{\rm f} (f_{\rm c}-f)\nonumber\\
              &  &+f_{\rm f}'(\rho_{\rm c})(\rho-\rho_{\rm c})\nonumber\\
              &  &+\frac{1}{2} f_{\rm f}''(\rho_{\rm c})
              (\rho-\rho_{\rm c})^{2} + \dots\;,
\end{eqnarray}
with $f_{\rm s}(\rho_{\rm c}) = f_{\rm f}(\rho_{\rm c})$, $f_{\rm
  s}'(\rho_{\rm c}) =
f_{\rm f}'(\rho_{\rm c})$, but $f_{\rm s}''(\rho_{\rm c}) > f_{\rm
  f}''(\rho_{\rm c})$ in
general.
Performing the inversion of $f(\rho)$ in order to get the density
profile leads to a piecewise linear
profile for the averaged density with two density jumps at $z=z_{1}$ and
$z=z_{2}$, as determined by the Maxwell construction, see
Fig.\ \ref{fig:sedimentation_profile_sketch}.
\begin{equation}
  \label{eq:density_profile_jumps}
  \rho(z) = \begin{cases}
    \frac{\mu-f_{\rm s}'(\rho_{\rm c})-mgz}{f_{\rm s}''(\rho_{\rm c})}
    + \rho_{\rm c}
    &\text{for $z_{1}<z<z_{2}$}\\
    \frac{\mu-f_{\rm f}'(\rho_{\rm c})-mgz}{f_{\rm f}''(\rho_{\rm
    c})} + \rho_{\rm c}
    &\text{else.}
  \end{cases}
\end{equation}
Consequently, by inserting this into the free energy function one gets
\begin{equation}
  \label{eq:Sigma_TD}
  \Sigma_{\rm TD} = -a (f - f_{\rm c}) l + \left(\frac{1}{f_{\rm
  f}''(\rho_{\rm c})} -
  \frac{1}{f_{\rm s}''(\rho_{\rm c})}\right) \frac{m^{2}g^{2}l^{3}}{12}.
\end{equation}
Note that $a>0$ in order to stabilize the solid for $f > f_{\rm c}$. 

Third, the elastic part can be calculated by continuum elastic
distortion theory of the solid. For 
a different situation of a solid in an 
external field this has been formulated by Gittes and Schick
\cite{Gittes:Schick:prb:30:209:1984}. Following these ideas, we assume 
a $z$-independent lateral strain $\epsilon_{\|}$ but consider a 
$z$-dependent vertical strain $\epsilon_{\bot}$. By symmetry,
$\epsilon_{\|}$ has to be zero for the crystal being stable at $z=z_0$,
i.e., for the crystal at the reentrant melting point. Elasticity theory 
predicts for $\Sigma_{\rm elast}$ 
\begin{equation}
  \label{eq:Sigma_elasticity}
  \Sigma_{\rm elast} \simeq 
  \frac{1}{2} \int_{-\frac{l}{2}}^{\frac{l}{2}} {\rm d}z\,
  C \epsilon_{\bot}^{2}(z),
 \end{equation}
where $C > 0$ is related to the elastic constants of the solid. As 
\begin{equation}
  \label{eq:epsilon_bot}
  \epsilon_{\bot} \propto \rho-\rho_{\rm c} \propto mgz,
\end{equation}
we obtain
\begin{equation}
  \label{eq:Sigma_elasticity_final}
  \Sigma_{\rm elast} = C' l^{3} m^{2} g^{2},
 \end{equation}
with another constant $C'$. Eq. (\ref{eq:Sigma_elasticity_final}) has a
similar form as the second term of Eq. (\ref{eq:Sigma_TD}).  

In summary, the total grand canonical excess free energy is
\begin{equation}
  \label{eq:excess_grand_canonical_free_energy_final}
  \Sigma_{\rm ex}(l) = -a(f-f_{\rm c})l + b\alpha^{2}l^{3} + \gamma_{1} + \gamma_{2},
\end{equation}
where $a,b > 0$. Moreover we will use $\gamma = \gamma_{1} +
\gamma_{2}$ from now on. A first order phase transition takes place if
the minimum of $\Sigma_{\rm ex}$ with respect to $l$ yields
$\Sigma_{\rm ex} = 0$.

This determines the following resulting scaling relations:
\begin{enumerate}
\item The realized crystalline thickness $l$ as obtained by minimizing 
  $\Sigma_{\rm ex}$ with respect to $l$ for fixed $\alpha$ and $f$ scales
  as
  \begin{equation}
      \label{eq:l_alpha_scaling}
      l \propto \frac{\sqrt{f-f_{\rm c}}}{\alpha}.
  \end{equation}
\item The phase transition to a crystalline sheet is first order. It
  happens beyond a critical $\alpha$-dependent arm number $f_{\rm crit}$
  where
  \begin{equation}
    \label{eq:f_alpha_scaling}
    f_{\rm crit}-f_{\rm c} \propto \gamma^{2/3} \alpha^{2/3},
  \end{equation}
  with a scaling exponent of $\frac{2}{3}$.
\item The width $l_{0}$ corresponding to the transition scales as 
  \begin{equation}
    \label{eq:l0_alpha_scaling}
    l_{0} \propto \gamma^{1/3} \alpha^{-2/3} \propto
    \frac{\gamma}{f-f_{\rm c}}.
  \end{equation}
\end{enumerate}
The analysis presented here is general, the scaling predictions
derived are valid for any reentrant melting behavior in equilibrium
(e.g., laser-induced freezing \cite{LIF1,LIF2} or polydisperse systems
\cite{Warren}).  
Furthermore all these relations can, in principle, be
checked by simulation. Relations (\ref{eq:f_alpha_scaling}) and
(\ref{eq:l0_alpha_scaling}), however, require high computational
efforts. In order to check on scaling relation
(\ref{eq:l_alpha_scaling}), we measured the crystal length $l$ in MC
simulations varying $\alpha$ or $f$, while keeping the density $\tau$
fixed. The crystal length is determined by the range $\Delta z$, where
the order parameter $\Psi_{4}(z)$ has values larger than $0.8$. The
results are plotted in Fig. \ref{fig:scaling}, showing excellent
agreement with the scaling predictions.

\begin{center}
\begin{figure}[hbt]
  \epsfig{file=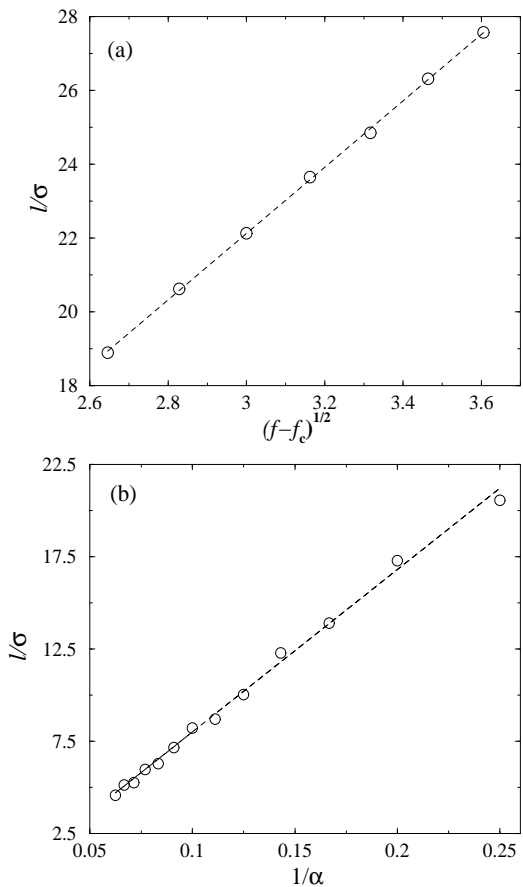, width=7cm, height=12cm, angle=0}
\caption{Verification of the scaling behavior
  theoretically predicted in Sec.\ \ref{sec:Phenomenological_Theory} in
  Eq.\ (\ref{eq:l_alpha_scaling}), calculated by MC simulations. In
  (a) the crystal length $l$ is plotted versus ${\sqrt{ f-f_{\rm
        c}}}$, keeping $\alpha=6.0$ fixed. In (b), $l$ is plotted
  versus $1/\alpha$ for a fixed arm number $f=39$. The dashed lines
  are linear fits to the simulation results (circles).}
\label{fig:scaling}
\end{figure}
\end{center}

\subsection{Weighted Density Approximation of the Density Functional}
\label{sec:WDA}
In order to verify the density oscillations close to the fluid-gas
interface of the sedimentation profiles for large values of $\alpha$,
we apply a simplified form of the HWDA (hybrid weighted density
approximation). The full HWDA was constructed by Leidl and Wagner in
\cite{leidl93}.  Given an external field $\Phi_{\rm ext}(z)$ the free
energy is a unique functional of the density profile $\rho(z)$. Thus,
the excess free energy per unit surface in the HWDA framework is
given by
\begin{eqnarray}
  {\cal F}_{\rm
    exc}[\rho]=\int_{0}^{\infty}\rho(z)f_{0}(\overline\rho(z)){\rm
    d}z,
\end{eqnarray}
where $f_{0}(\rho)$ denotes the excess free energy per particle of a
homogeneous liquid of density $\rho$. The weighted density
$\overline\rho(z)$ follows from a convolution with the weighting
function $\omega(r;\rho)$
\begin{eqnarray}
  \overline\rho(z)=\int\rho({z'})\omega(|{\bf
    r-r'}|;\hat\rho){\rm d}{\bf r'}
\end{eqnarray}
with a global density $\hat\rho$. The weighting function $\omega
(r;\rho)$ is fixed by a simple quadratic equation in Fourier space \cite{leidl93}:
\begin{eqnarray}
  2f_{0}'(\rho_{0})\tilde\omega(k;\rho_{0}) 
  &+ &\rho_{0}f_{0}''(\rho_{0})\tilde\omega^{2}(k;\rho_{0})\nonumber\\
  &= &-\beta^{-1}\tilde c^{(2)}(k;\rho_{0})
\end{eqnarray}
The primes denote differentiations with respect to the density $\rho$
and $\tilde c^{(2)}(k;\rho_{0})$ is the Fourier transform of the
direct correlation function of the homogeneous fluid. A unique
solution of $\tilde\omega(k;\rho_{0})$ is determined by the
normalization $\tilde \omega(k=0;\rho_{0})=1$, also ensuring the
compressibility rule to be satisfied. We have solved the homogeneous
problem with Ornstein-Zernike fluid integral equations using the
Rogers-Young closure \cite{RY}. Resulting correlation functions and
structure factors are in very good agreement with MC simulations of
the bulk system \cite{watz_struc98}.  In difference to the complete
HWDA, where the global density $\hat\rho$ is chosen to be a functional
of $\rho(z)$, we keep $\hat\rho$ fixed. This simplification is
sufficient to verify the observed oscillations, accompanied by the
advantage that the numerical effort is enormously reduced. Best
agreement with simulation results could be achieved when choosing the
global density $\hat\rho$ to be of the order of the averaged density
near the bottom wall $z=0$. The tails of the density profiles are
nearly unaffected 
by the choice of $\hat\rho$. A similar approach is used in the SWDA
\cite{kim96} for an inhomogeneous fluid in contact with a bulk fluid
of density $\rho_{b}$; there $\hat\rho$ was chosen to be $\rho_{b}$.

Applying the usual Euler-Lagrange minimization for the Helmholtz free
energy per unit area $\cal F[\rho]$ with chemical potential $\mu$
\begin{eqnarray}
  \frac{\delta \cal F}{\delta \rho(z)}=\mu-\Phi_{\rm ext}(z),
\end{eqnarray}
and using $\Phi_{\rm ext}(z)=V_{\rm sw}(z)+\alpha z/\beta\sigma$,
 we obtain for the density profile $\rho(z)$
\begin{eqnarray}
  \label{rho}
  \rho(z) = \begin{cases} 
    \xi\exp\{c^{(1)}(z;[\rho])-\alpha
    z/\sigma-\beta V_{\rm sw}(z)\} &z>0\\
    \cr 0 &\text{else}.
  \end{cases}
\end{eqnarray}
The fugacity $\xi$ is determined by the normalization condition
$\xi=\tau/\int_{0}^{\infty}{\rm d}z\exp\{c^{(1)}(z;[\rho])-\alpha
z/\sigma-\beta V_{\rm sw}(z)\}$. $c^{(1)}(z;[\rho])$ is the one-particle
correlation function:
\begin{eqnarray}
  -\beta^{-1}c^{(1)}(z;[\rho]) &= &\frac{\delta \cal F_{\rm ex}[\rho]}{\delta \rho(z)}\\
                    &= &f_{0}(\overline\rho(z))\nonumber\\
                    & &+\!\!\int {\rm d}z' \rho(z')
                    f'_{0}(\overline\rho(z'))\omega(|z-z'|;\hat\rho).\nonumber
\end{eqnarray}
Equation (\ref{rho}) was solved for the profile by standard iterative
techniques, see, e.g., Ref.\ \cite{denton91}. The results for an arm number
$f=32$, $\tau\sigma^{2}=21.8$ and three different values of $\alpha$ are shown in
Fig. \ref{fig:dft} together with MC simulation data.
\begin{figure}[hbt]
 \epsfig{file=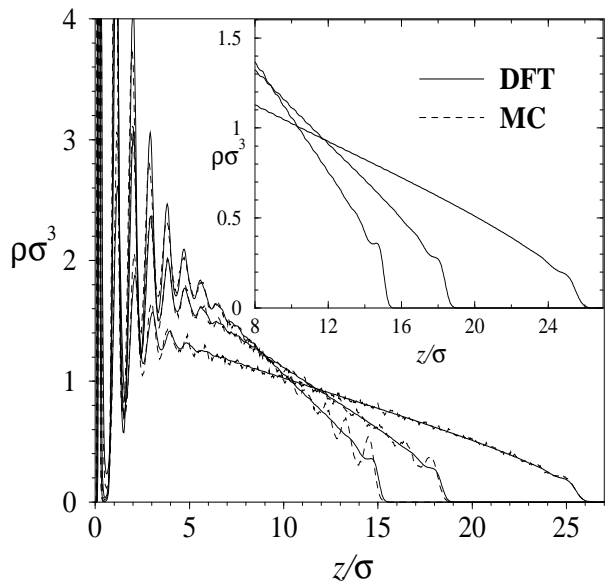,width=8cm,height=8cm,angle=0}  
  \caption{Density profiles $\rho(z)\sigma^{3}$ for three different
    values of $\alpha=10.0,20.0,30.0$ and fixed arm number
    $f=32$ and fixed density $\tau\sigma^{2}=21.8$ calculated with DFT (solid curves)
    compared to MC simulation results (dashed curves). The slope of the
    curves increases with increasing $\alpha$. The inset shows
    the bare DFT results for a better identification of the interface
    oscillations.} 
  \label{fig:dft}
\end{figure}
The global density $\hat\rho$ for all three profiles is fixed at
$\hat\rho\sigma^{3}=1.8$. The DFT results are in very good agreement
with the simulation profiles. In particular, the interface
oscillations with wavelength $\sigma$ also occur in the DFT. 
For $\alpha=10.0$, the lowest $\alpha$ that is shown, the profile
is nearly indistinguishable from MC data far from the wall, while for
increased $\alpha$, $\alpha=20.0$ and $\alpha=30.0$ the interface
oscillations are underestimated in our DFT approach.

These oscillations are not a specific feature of the star polymers;
we have also performed MC simulations using a repulsive Yukawa interaction 
of the form 
\begin{eqnarray}
V(r)\propto \exp(-\kappa r)/r,
\end{eqnarray}
corresponding to the part of the star polymer pair potential 
(\ref{eq:star_star_potential}) valid for distances $r\geq \sigma$.
Here exactly the same behavior could 
be found at the tails of the density profiles.
For hard sphere systems, on the other hand,
no such density oscillations are present. This suprising fact might be  
attributed to the long ranged tail in the interaction potential.
 
\section{Conclusions}
\label{sec:Conclusion}
Concluding, we have presented results for systems exhibiting reentrant
melting in the bulk phase diagram, under gravitational influence. It
was shown that a phase transition occurs when the gravitational
strength $\alpha$ is varied: Below a critical $\alpha^{*}(f, \tau)$,
intercalated crystallization occurs in the sedimentation profiles of
the observed star-polymer solutions, whereas for $\alpha >
\alpha^{*}(f, \tau)$ we find monotonic sedimentation profiles
$\rho(z)$. In MC computer simulations scaling relations for the
crystallization, predicted in the framework of a phenomenological
theory, valid for all systems exhibiting reentrant melting in the bulk
phase diagram, could be verified. Using density functional theory,
density oscillations at the fluid-gas boundary, observed in the MC
simulations, could be reproduced. 

In principle, our results can be verified in surface-sensitive scattering
experiments or real-space imaging methods for colloidal suspensions.
Unlike non-monotonicities on the liquid side of the gas-liquid
equilibrium interface (see e.g. \cite{inter}),
the density oscillations on top of the
sedimentation profile are not affected by capillary wave fluctuations
and may thus be verified in real samples. The intervening solid sheet
should be signalled by a Bragg-like peak in surface reflection
measurements.

\acknowledgments

The authors wish to thank M.\ Heni, M.\ Schmidt, and A.\ Esztermann for
stimulating discussions and A.\ Esztermann for a critical reading of
the manuscript. This work has been supported by the 
Deutsche Forschungsgemeinschaft within the SFB 237.

\end{document}